\documentclass{article}%
\usepackage{amsfonts}
\usepackage{amssymb}
\usepackage{amsmath}
\usepackage{graphicx}%
\providecommand{\U}[1]{\protect\rule{.1in}{.1in}}
\begin{document}

\title{Born Again! The Born Rule as a feature of superposition}
\author{David Ellerman\\Independent Researcher, Ljubljana, Slovenia\\david@ellerman.org\\https://orcid.org/0000-0002-5718-618X}
\maketitle

\begin{abstract}
Where does the Born Rule come from? We ask: ``What is the simplest extension of
probability theory where the Born rule appears''? This is answered by introducing ``superposition events'' in addition to the usual discrete events into ordinary probability theory. Two-dimensional matrices (e.g., incidence matrices and density matrices) are needed to mathematically represent the differences between the two types of events. Then it is shown that those incidence and density matrices for superposition events are the (outer) products of a vector and its transpose (whose components foreshadow the ``amplitudes'' of quatnum mechanics). The squares of the components of those ``amplitude'' vectors yield the probabilities of the outcomes. That is how the Born Rule arises in the natural mathematical extension of probability theory to include superposition events formulated over the reals--all of which is naturally extended to the full Born Rule in the Hilbert spaces over the complex numbers of quantum mechanics.


\end{abstract}
\tableofcontents

\section{Introduction}

In quantum mechanics (QM), the Born Rule provides the all-important link
between the mathematical formalism (e.g., the wave function) and experimental
results in terms of probabilities. The rule does not occur in ordinary
classical probability theory. Hence one might ask with Steven Weinberg:
\textquotedblleft So where does the Born Rule come from?\textquotedblright%
\ \cite[p. 92]{weinberg:qm} Can it be derived from the other postulates of QM
or must it be assumed as an additional postulate? There is a vast and
sophisticated literature debating these questions--see \cite{masanes:born},
\cite{vaidman:born}, \cite{hossenfelder:born}, and the articles cited therein.

In this paper, a different approach is taken. What is the simplest extension
to classical probability theory where the Born Rule appears? We expand
ordinary finite probability theory by introducing superposition events in
addition to the usual discrete events (subsets of the outcome space) and then
we show that the Born Rule naturally arises in the mathematics of
superposition events. A superposition event is a purely mathematical notion in
this enriched probability theory--although obviously inspired by the notion of
a superposition state in quantum mechanics.

It is not a coincidence that superposition (including the special case of
entanglement) is the key non-classical notion in quantum mechanics.

\begin{itemize}
\item For instance, \textquotedblleft\textit{superposition}, with the
attendant riddles of entanglement and reduction, remains\textit{ the }central
and generic interpretative problem of quantum theory.\textquotedblright%
\ \cite[p. 27]{cushing:qft} 

\item Some writers use the word \textquotedblleft
entanglement\textquotedblright\ to mean or include superposition.\footnote{The
argument by some that it is only entanglement proper that is characteristic of
QM, since there is superposition in classical electromagnetic waves or in
water waves, will be addressed below.} \textquotedblleft The superposition or
`entanglement' of states is a hallmark of quantum mechanics.\textquotedblright%
\ \cite[p. 50]{bunge:m-and-m}\ 

\item \textquotedblleft In this sense, one can say that the entanglement
arising from summation of probability amplitudes over all possible Feynman
paths in the appropriate configuration space is \textit{the} distinctive
feature of quantum mechanics, the sole mystery.\textquotedblright\ \cite[p.
248]{stachel:feynman-paths} 
\end{itemize}

Dirac was quite clear on this point from the beginning.

\begin{quotation}
The nature of the relationships which the superposition principle requires to
exist between the states of any system is of a kind that cannot be explained
in terms of familiar physical concepts. One cannot in the classical sense
picture a system being partly in each of two states and see the equivalence of
this to the system being completely in some other state. There is an entirely
new idea involved, to which one must get accustomed and in terms of which one
must proceed to build up an exact mathematical theory, without having any
detailed classical picture. \cite[p. 12]{dirac:principles}
\end{quotation}

\noindent As a purely mathematical notion (as developed here), superposition
events could have been (but were not) introduced long before QM. The thesis is
that the Born Rule is not a bug that needs to be \textquotedblleft
explained\textquotedblright\ or \textquotedblleft justified\textquotedblright;
it is just a feature of the vectorial mathematics of a superposition event
foreshadowed in this minimally expanded probability theory--and then extended
to the Hilbert spaces over $\mathbb{C}$ in QM.

\section{Intuitively modeling superposition events}

In classical finite probability theory, the \textit{outcome} (or
\textit{sample}) \textit{space} is a set $U=\{  u_{1},...,u_{n}\}
$ (where we assume equal probabilities until different point probabilities are
introduced). An (ordinary) \textit{event} $S$ is a non-empty subset
$S\subseteq U$. In an (ordinary) event $S$, the atomic outcomes or elements of
$S$ are considered as perfectly discrete and distinguished from each other; in
each run of the \textquotedblleft experiment\textquotedblright\ or trial,
there is the probability $\Pr(  S)  $ occurring and the probability
$\Pr(  T|S)  $ of an event $T\subseteq U$ occurring given the $S$
occurs (including the case of a specific outcome $T=\{  u_{i}\}  $).

The intuitive idea of the corresponding superposition state, denoted $\Sigma
S$, is that the outcomes in the state are not distinguished from each other
but are blobbed or cohered together as an indefinite event. The concept of
superposition includes the notion of \textit{amplitudes} as the relative
`strength' of the outcomes in the superposition. It is the rules for dealing with amplitudes
that separates quantum probabilities from ordinary probabilities.

\begin{quotation}
\noindent\ In the two-slit experiment, for example, passage through one slit
or the other is only a distinguishable alternative if a counter is placed
behind one of the slits; without such a counter, these are indistinguishable
alternatives. Classical probability rules apply to distinguishable processes.
Nonclassical probability amplitude rules apply to indistinguishable processes.
\cite[p. 314]{stachel:needqlogic}
\end{quotation}

\noindent Hence we are considering the minimal extension of classical
probability theory that includes superposition events and \textit{thus} also
the notion of amplitudes in the superposition. No physics is involved in this
extension; we are only investigating what emerges naturally from the
\textit{mathematics} with these concepts introduced into otherwise classical
probability theory.

In each run of the \textquotedblleft experiment\textquotedblright\ or trial
conditioned on $\Sigma S$, the indefinite event is sharpened to a less
indefinite event which is maximally sharpened to one of the definite outcomes
in $S$. The probabilities of the individual outcomes are assumed to be the same when conditioned by the discrete event or the superposition event: $\operatorname*{Pr}(u_{i}\vert S)= \operatorname*{Pr}(u_{i}\vert \Sigma S)=p_{i}$ where $p=(p_{1},...,p_{n})$ are the point probabilities.\footnote{This is not a bug but a feature since in QM, the probabilities of the eigenstates in a superposition are the same as the probabilities in the corresponding completely decomposed mixed state.} In the case of a singleton event $S=\{  u_{i}\}  $, the
ordinary event $S=\{  u_{i}\}  $ is the same as the superposition
event $\Sigma S=\Sigma\{  u_{i}\}  =\{  u_{i}\}  =S$.

For a suggestive visual example, consider the outcome set $U$ as a pair of
isosceles triangles that are distinct by the labels on the equal sides and the
opposing angles.%

\begin{center}
\includegraphics[
height=0.9608in,
width=2.4907in
]%
{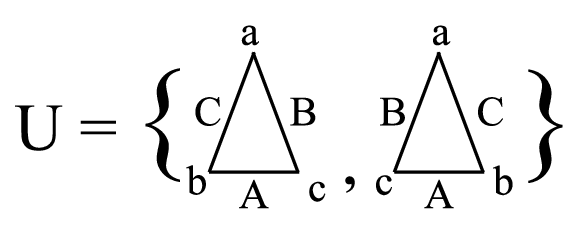}%
\end{center}

\begin{center}
Figure 1: Set of distinct isosceles triangles
\end{center}

\noindent The superposition event $\Sigma U$ is definite on the properties
that are common to the elements of $U$, i.e., the angle $a$ and the opposing
side $A$, but is indefinite where the two triangles are distinct, i.e., the
two equal sides and their opposing angles are not distinguished by labels
(\cite{ell:ftm}; \cite{ell:3theories}).

\begin{center}%

\begin{center}
\includegraphics[
height=0.896in,
width=3.103in
]%
{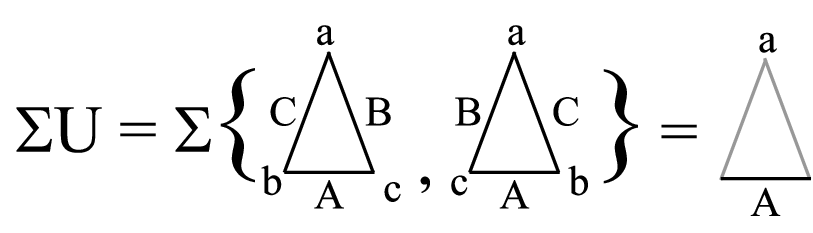}%
\end{center}

Figure 2: The superposition event $\Sigma U$.
\end{center}

For another visual example of superposition, consider a square with labeled
corners $\fbox{$_d^a\,_c^b$}$. If it is flipped around the diagonal, then it
becomes $\fbox{$_b^a\,_c^d$}$ and then the superposition is $\sum(
\fbox{$_d^a\,_c^b$},\fbox{$_b^a\,_c^d$})  =\fbox{$_?^a\,_c^?$}$. This
sort of simple superposition (without magnitudes) is indefinite on where the
superposed entities differ and definite on where they are the same.

It might be noted that this notion of superposition and the notion of
abstraction are essentially flip-side viewpoints of the same idea of
extracting from a set an entity that is definite on the commonalities of the
elements of the set and indefinite (or silent) on where the elements differ
\cite{ell:abstr}. The two flip-side viewpoints are like seeing a glass
half-empty (superposition) or seeing a glass half-full (abstraction).

\section{Mathematically modeling superposition events}

What is a mathematical model that will distinguish between the ordinary discrete event
$S$ and the superposition event $\Sigma S$? Using $n$-ary column vectors in
$\mathbb{R}^{n}$, the ordinary event $S$ could be represented by the column
vector, denoted $\vert S\rangle $, with the $i^{th}$ entry
$\chi_{S}(  u_{i})  $, where $\chi_{S}:U\rightarrow\{
0,1\}  $ is the characteristic function for $S$, i.e., $\chi_{S}%
(u_{i})=1$ if $u_{i}\in S$, else $0$. That vector representation is \textit{insufficient} to represent whether the elements of $S$ are superposed or not. Hence to represent the superposition event
$\Sigma S$ we need to add a dimension to use two-dimensional $n\times n$
matrices to represent the blobbing together or cohering of the elements of $S$
in the superposition event $\Sigma S$.

An \textit{incidence matrix} for a binary relation $R\subseteq U\times U$ is
the $n\times n$ matrix $\operatorname*{In}(  R)  $ where
$\operatorname*{In}(  R)  _{jk}=1$ if $(  u_{j},u_{k})
\in R$, else $0$. The diagonal $\Delta S$ is the binary relation consisting of
the ordered pairs $\{  (  u_{i},u_{i})  :u_{i}\in S\}  $
and its incidence matrix $\operatorname*{In}(  \Delta S)  $ is the
diagonal matrix with the diagonal elements $\chi_{S}(  u_{i})  $.
The superposition state $\Sigma S$ could then be represented as
$\operatorname*{In}(  S\times S)  $, the incidence matrix of the
binary relation $S\times S\subseteq U\times U$, where the non-zero
off-diagonal elements represent the equating, cohering, or blobbing together
of the corresponding diagonal elements.\footnote{On the universe set $U$, the
binary relation $U\times U$ is the universal equivalence relation which
equates all the elements of $U$. Thus $S\times S$ is the universal equivalence
relation on $S$ which equates all its elements.}

Given two column vectors $\vert s\rangle =(  s_{1}
,...,s_{n})^{t}$ and $\vert t\rangle =(
t_{1},...,t_{n})  ^{t}$ in $\mathbb{R}^{n}$(where $(  {})  ^{t}$ is the transpose), their \textit{inner product} is the sum of the products of the corresponding entries and is
denoted $\langle t|s\rangle =(  \vert t\rangle
)  ^{t}\vert s\rangle =\sum_{i=1}^{n}t_{i}s_{i}$.  A vector $\vert s\rangle
$ is \textit{normalized} if $\langle s|s\rangle =1$. 

To explain ``where the Born Rule comes from,'' we have to also show how probability \textit{amplitudes} arise in our extension of classical probability theory. This is foreshadowed even at the level of incidence matrices. The \textit{outer product} $\vert
s\rangle \langle t\vert$ is the $n\times n$ matrix denoted as $\vert
s\rangle \langle t\vert =$ $\vert s\rangle (
\vert t\rangle )  ^{t}$. The key result, that foreshadows probability amplitudes, is that the outer product of $\vert S \rangle$ with its transpose is the incidence matrix  representing the superposition event.
\begin{center}
	$\vert S \rangle \langle S \vert = \operatorname*{In}(S\times S)$.
\end{center} 
The proof is simply that $\chi_{S}(u_{i})\chi_{S}(u_{j}) =1$ if and only if (iff) $(u_{i},u_{j})\in S\times S$. This is a key result because it directly connects the outer product of a vector with its transpose with superposition. The vector in the outer product foreshadows the vector of ``amplitudes'' that is used in the Born Rule and thereby the connection with superposition.

If we divided $\operatorname*{In}(  \Delta S)  $ and
$\operatorname*{In}(  S\times S)  $ through by their trace (sum of
diagonal elements) $\vert S\vert $, then we obtain two density
matrices $\rho(  S)  =\frac{\operatorname*{In}(  \Delta
S)  }{\vert S\vert }$ and: 
\begin{center}
	$\rho(  \Sigma S)
	=\frac{\operatorname*{In}(  S\times S)  }{\vert S\vert
	}=\frac{1}{\sqrt{\vert S\vert }}\vert S\rangle
	\langle S\vert \frac{1}{\sqrt{\vert S\vert }}$ 
\end{center}
\noindent over the reals $\mathbb{R}$. In the case of equiprobable outcomes $p_{i}= \frac{1}{n}$, we already have a special case of the Born Rule for the probability of $u_{i}$ given the superposition event $\Sigma S$:
\begin{center}
	$\langle u_{i} \vert \frac{1}{\sqrt{\vert S}\vert} S \rangle ^{2}=\frac{1}{\vert S \vert}\chi_{S}(u_{i})$.
\end{center}

In general, a \textit{density matrix} $\rho$ over the reals $\mathbb{R}$ (or the complex numbers $\mathbb{C}$) is a symmetric matrix $\rho=\rho^{t}$ (or conjugate symmetric matrix
$\rho=(  \rho^{\ast})  ^{t}$ in the case of $\mathbb{C}$) with trace $\operatorname*{tr}[  \rho]  =1$ and all non-negative
eigenvalues which sum to $1$. 

The analogue to a probability theory discrete event in QM is a completely discrete (or decomposed) mixed state. It is not a vector in Hilbert space. A vector in Hilbert space represents a pure state which is in general a superposition in a given basis and thus it is the analogue of a superposition event. One virtue of density matrices is that they represent both mixed and pure states. 

A density matrix $\rho$ is \textit{pure} if
$\rho^{2}=\rho$, otherwise a \textit{mixture}. The existence of the non-zero
off-diagonal elements in the incidence matrices and thus in the density
matrices indicates the presence of not only superposition but also amplitudes
indicating the coherence of the superposed outcomes.

\begin{quotation}
\noindent For this reason, the off-diagonal terms of a density matrix ... are
often called \textit{quantum coherences} because they are responsible for the
interference effects typical of quantum mechanics that are absent in classical
dynamics. \cite[p. 177]{auletta:qm}
\end{quotation}

Consider the partition $\pi=\{  B_{1},B_{2}\}  =\{  \{
\diamondsuit,\heartsuit\}  ,\{  \clubsuit,\spadesuit\}
\}  $ on the outcome set $U=\{  \clubsuit,\diamondsuit
,\heartsuit,\spadesuit\}  $ with equiprobable outcomes like drawing
cards from a randomized deck. For instance, the superposition event associated
with $B_{1}=\{  \diamondsuit,\heartsuit\}  $, has a pure density
matrix since (rows and columns labeled in the order $\{  \clubsuit
,\diamondsuit,\heartsuit,\spadesuit\}  $):

\begin{center}
$\rho(  \Sigma B_{1})  =\frac{1}{\sqrt{\vert B_{1}\vert
}}%
\begin{bmatrix}
0\\
1\\
1\\
0
\end{bmatrix}%
\begin{bmatrix}
0 & 1 & 1 & 0
\end{bmatrix}
\frac{1}{\sqrt{\vert B_{1}\vert }}=\frac{1}{\vert
B_{1}\vert }%
\begin{bmatrix}
0 & 0 & 0 & 0\\
0 & 1 & 1 & 0\\
0 & 1 & 1 & 0\\
0 & 0 & 0 & 0
\end{bmatrix}
=%
\begin{bmatrix}
0 & 0 & 0 & 0\\
0 & \frac{1}{2} & \frac{1}{2} & 0\\
0 & \frac{1}{2} & \frac{1}{2} & 0\\
0 & 0 & 0 & 0
\end{bmatrix}
$
\end{center}

\noindent equals its square, but density matrix for the discrete set $B_{1}$:

\begin{center}

$\rho(  B_{1})  =%
\begin{bmatrix}
0 & 0 & 0 & 0\\
0 & \frac{1}{2} & 0 & 0\\
0 & 0 & \frac{1}{2} & 0\\
0 & 0 & 0 & 0
\end{bmatrix}
$
\end{center}

\noindent is a mixture since it does not equal its square.

Intuitively, the interpretation of the superposition event represented by
$\rho(  \Sigma B_{1})  =\rho(  \Sigma\{  \diamondsuit
,\heartsuit\}  )  $ is that it is definite on the properties
common to its elements, e.g., in this case, being a red suite, but indefinite
on where the elements differ. The indefiniteness is indicated by the non-zero
off-diagonal elements that indicate that the diamond suite $\diamondsuit$ is
blurred, cohered, or superposed with the hearts suite $\heartsuit$ in the
superposition state $\Sigma\{  \diamondsuit,\heartsuit\}  $.

The next step is to bring in general point probabilities $p=(  p_{1}%
,...,p_{n})  $ where those two real density matrices $\rho(
S)  $ and $\rho(  \Sigma S)  $ defined above correspond to
the special case of the equiprobable distribution on $S$ with $0$
probabilities outside of $S$.

\section{Density matrices with general probability distributions}

Let the outcome space $U=\{  u_{1},...,u_{n}\}  $ have the strictly
positive probabilities $p=\{  p_{1},...,p_{n}\}  $. The probability
of a (discrete) subset $S$ is $\Pr(  S)  =\sum_{u_{i}\in S}p_{i}$
and the conditional probability of $T\subseteq U$ given $S$ is: $\Pr(
T|S)  =\frac{\Pr(  T\cap S)  }{\Pr(  S)  }$. But
we have now reformulated both the usual discrete event $S$ and the new
superposition event $\Sigma S$ in matrix terms. Hence we need to reformulate
the usual conditional probability calculation in matrix terms and then apply
the same matrix operations to define the conditional probabilities for the
superposition events.

The density matrix $\rho(  U)  $ is the diagonal matrix with the
point probabilities down the diagonal. Let $P_{S}$ be the diagonal
(projection) matrix with the diagonal entries $\chi_{S}(  u_{i})
$. Then $\Pr(  S)  $ can be computed by replacing the summation
$\sum_{u_{i}\in S}p_{i}$ with the trace formula: $\Pr(  S)
=\operatorname*{tr}[P_{S}\rho(  U)  ]$. The density matrix
$\rho(  S)  $ for the classical discrete $S$ is defined as the
diagonal matrix with diagonal entries $\frac{p_{i}}{\Pr(  S)  }$ if
$u_{i}\in S$, else $0$, which yields the mixture density matrix $\rho(
S)  $. For $\rho(  S)  $, the eigenvalues are just the
conditional probabilities $\Pr(  \{  u_{i}\}  |S)
=\frac{\Pr(  \{  u_{i}\}  \cap S)  }{\Pr(
S)  }=\frac{p_{i}}{\Pr(  S)  }\chi_{S}(  u_{i})  $
for $i=1,...,n$. Then the conditional probability $\Pr(  T|S)  $ is
reproduced in the matrix format as:

\begin{center}
$\Pr(  T|S)  =\operatorname*{tr}[  P_{T}\rho(  S)
]  $.
\end{center}

The previously constructed density matrix $\rho(  \Sigma S)
=\frac{1}{\sqrt{\vert S\vert }}\vert S\rangle
\langle S\vert \frac{1}{\sqrt{\vert S\vert }}$ for the
superposition event $\Sigma S$ was for the special case of equiprobable
outcomes. In the general case of point probabilities, the column vector
$\frac{1}{\sqrt{\vert S\vert }}\vert S\rangle $ is
generalized to $\vert s\rangle $ where the $i^{th}$ entry,
symbolized $\langle u_{i}|s\rangle $, is the \textit{amplitude}
$\sqrt{\frac{p_{i}}{\Pr(  S)  }}$ if $u_{i}\in S$, else $0$, and then

\begin{center}
$\rho(  \Sigma S)  =\vert s\rangle \langle
s\vert $
\end{center}

\noindent which is a pure state density matrix. For the pure density matrix
$\rho(  \Sigma S)  $, there is one eigenvalue of $1$ with the rest
of the eigenvalues being zeros (since the sum of the eigenvalues is the
trace). Given just $\rho(  \Sigma S)  $, the vector $\vert
s\rangle $ is recovered as the normalized eigenvector associated with
the eigenvalue of $1$ and $\rho(  \Sigma S)  =\vert
s\rangle \langle s\vert $.\footnote{This is by the spectral
decomposition of that real density matrix as a Hermitian operator.} Since the ``amplitude'' vector $\vert s \rangle$ arises from the representation of $\rho(\Sigma S)$ as an outer product (foreshadowed even for incidence matrices) and since the special case of the Born Rule holds in this extended probability theory:
\begin{center}
	$\langle u_{i} \vert s \rangle ^{2}=\frac{p_{i}}{\operatorname*{Pr}(S)}\chi_{S}(u_{i})$
\end{center}
\noindent we already see how the Born Rule is a feature of the mathematics of superposition.

The probabilities computed for the classical and superposition events will be
the same--which is a feature, not a bug, since the same thing occurs in
quantum mechanics.\footnote{For instance, a spin measurement along, say, the
$z$-axis of an electron cannot distinguish between the superposition state
$\frac{1}{\sqrt{2}}(  \vert \uparrow\rangle +\vert
\downarrow\rangle )  $ with a density matrix like $\rho(
\Sigma U)  $ and a statistical mixture of half electrons with spin up
and half with spin down with a density matrix like $\rho(  U)  $
\cite[p. 176]{auletta:qm}.} It is the interpretation, not the probabilities,
that are different for the two types of events. For discrete events, the given
discrete event $S$ is reduced by conditioning to the discrete event $T\cap S$.
For superposition events, the given superposition event $\Sigma S$ is
sharpened (i.e., made less indefinite) to the superposition event
$\Sigma(  T\cap S)  $ with the probability $\Pr(
\Sigma(  T\cap S)  |\Sigma S)  =\Pr(  \Sigma T|\Sigma
S)  $ given the event $\Sigma S$.

A \textit{partition} $\pi=\{  B_{1},...,B_{m}\}  $ on $U$ is a set
of non-empty subsets, called \textit{blocks}, $B_{j}\subseteq U$ that are
disjoint and whose union is $U$. Taking each block $B_{j}=S$, then there is
the normalized column vector $\vert b_{j}\rangle $ whose $i^{th}$
entry is $\sqrt{\frac{p_{i}}{\Pr(  B_{j})  }}\chi_{B_{j}}(
u_{i})  $ and the density matrix $\rho(  \Sigma B_{j})
=\vert b_{j}\rangle \langle b_{j}\vert $ for the
superposition subset $\Sigma B_{j}$. Then the density matrix $\rho(
\pi)  $ for the partition $\pi$ is just the probability sum of those
pure density matrices for the superposition blocks:

\begin{center}
$\rho(  \pi)  =\sum_{j=1}^{m}\Pr(  B_{j})  \rho(
\Sigma B_{j})  $.
\end{center}

\noindent The eigenvalues for $\rho(  \pi)  $ are the $m$
probabilities $\Pr(  B_{j})  $ with the remaining $n-m$ values of
$0$.

Given two partitions $\pi=\{  B_{1},...,B_{m}\}  $ and
$\sigma=\{  C_{1},...,C_{m^{\prime}}\}  $, the partition $\pi$
\textit{refines} the partition $\sigma$, written $\sigma\precsim\pi$, if for
each block $B_{j}\in\pi$, there is a block $C_{j^{\prime}}\in\sigma$ such that
$B_{j}\subseteq C_{j^{\prime}}$. The partitions on $U$ form a partial order
under refinement. The maximum partition or top of the order is the
\textit{discrete partition} $\mathbf{1}_{U}=\{  \{  u_{i}\}
\}  _{i=1}^{n}$ where all the blocks are singletons and the minimum
partition or bottom is the \textit{indiscrete partition} $\mathbf{0}
_{U}=\{  U\}  $ with only one block $U$ where all the elements of $U$ are blobbed together. Then the density matrices
for these top and bottom partitions are just the density matrices for the
discrete set $U$ and the superposition set $\Sigma U$:

\begin{center}
$\rho(  \mathbf{1}_{U})  =\rho(  U)  $ and $\rho(
\mathbf{0}_{U})  =\rho(  \Sigma U)  $.
\end{center}

\noindent The same holds if we cut down to any event $S\subseteq U$, i.e., $\rho(  \mathbf{1}_{S})  =\rho(  S)  $ and $\rho(
\mathbf{0}_{S})  =\rho(  \Sigma S)=\vert s\rangle \langle
s\vert  $. Since $\mathbf{0}_{S}$ represents the blobbing together of the elements of $S$ and $\mathbf{1}_{S}$ represents the discrete set $S$, i.e., the event $S$ in ordinary finite probability theory, this result comports completely with the mathematical treatment of superposition events $\Sigma S$ as opposed to discrete events $S$.

Let us illustrate this result with the case of flipping a fair coin. The
classical set of outcomes $U=\{  H,T\}  $ is represented by the
density matrix:

\begin{center}
$\rho(  U)  =$ $%
\begin{bmatrix}
\frac{1}{2} & 0\\
0 & \frac{1}{2}%
\end{bmatrix}
$.

\begin{center}
\includegraphics[
height=1.0836in,
width=3.7239in
]%
{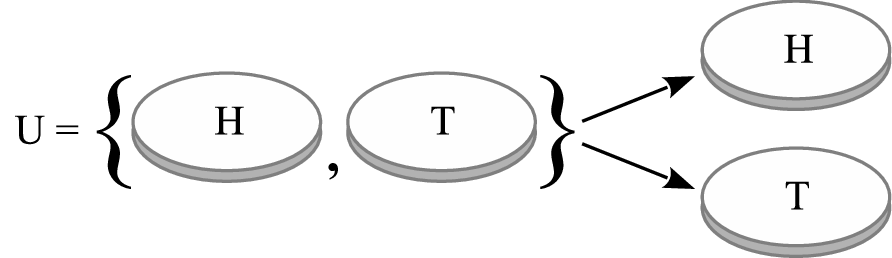}%
\end{center}

Figure 3: Classical event: A trial picks out heads or tails.
\end{center}

\noindent The superposition event $\Sigma U$, that blends or superposes heads
and tails, is represented by the density matrix:

\begin{center}
$\rho(  \Sigma U)  =%
\begin{bmatrix}
\frac{1}{2} & \frac{1}{2}\\
\frac{1}{2} & \frac{1}{2}%
\end{bmatrix}
$.%

\begin{center}
\includegraphics[
height=1.2773in,
width=2.687in
]%
{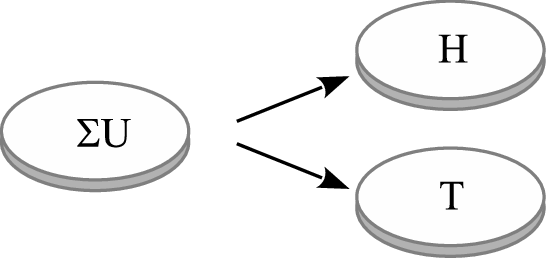}%
\end{center}

Figure 4: Superposition event: A trial sharpens to heads or tails.
\end{center}

\noindent The probability of getting heads in each case is:

\begin{center}
$\Pr(  H|\rho(  U)  )  =\operatorname{tr}[
P_{\{  H\}  }\rho(  U)  ]  =\operatorname{tr}%
[
\begin{bmatrix}
1 & 0\\
0 & 0
\end{bmatrix}%
\begin{bmatrix}
\frac{1}{2} & 0\\
0 & \frac{1}{2}%
\end{bmatrix}
]  =\operatorname{tr}%
\begin{bmatrix}
\frac{1}{2} & 0\\
0 & 0
\end{bmatrix}
=\frac{1}{2}$

$\Pr(  H|\rho(  \Sigma U)  )  =\operatorname{tr}[
P_{\{  H\}  }\rho(  \Sigma U)  ]
=\operatorname{tr}[
\begin{bmatrix}
1 & 0\\
0 & 0
\end{bmatrix}
\begin{bmatrix}
\frac{1}{2} & \frac{1}{2}\\
\frac{1}{2} & \frac{1}{2}%
\end{bmatrix}
]  =\operatorname{tr}\allowbreak%
\begin{bmatrix}
\frac{1}{2} & \frac{1}{2}\\
0 & 0
\end{bmatrix}
=\frac{1}{2}$
\end{center}

\noindent and similarly for tails. Thus the two conditioning events $U$ and
$\Sigma U$ cannot be distinguished by performing an experiment or trial that
distinguishes heads and tails. As noted, this is a feature, not a bug, since
the same thing occurs in quantum mechanics. In QM, they can only be
distinguished by measurement in a different observable basis (see
\cite{ell:abstr} for an example).

It might be noted that the density matrix $\rho(\mathbf{1}_{U})$ for the
discrete partition $\mathbf{1}_{U}$ is the density matrix $\rho(
U)  $ for the classical discrete set $U$ which is like a
\textquotedblleft statistical mixture describing the state of a classical dice
before the outcome of the throw.\textquotedblright\ \cite[p. 176]{auletta:qm}
In the logic of partitions (or equivalence relations) \cite{ell:lop2apps} and
its quantitative version, logical information theory based on logical entropy
\cite{ell:nf4it}, the \textit{distinctions} or dits of a partition
$\pi=(  B_{1},...,B_{m})  $ on $U$ are the ordered pairs in
$U\times U$ whose elements are in different blocks of the partition. The set of all
distinctions is the ditset $\operatorname*{dit}(  \pi)
=\cup_{j,k=1;j\neq k}^{m}B_{j}\times B_{k}$. The complementary set in $U\times
U$ is the set of all \textit{indistinctions} or \textit{indits} (ordered pairs
in the same block) is $\operatorname*{indit}(  \pi)  =\cup
_{j=1}^{m}B_{j}\times B_{j}$ which is the equivalence relation associated with
the partition $\pi$. The classical nature of the discrete partition
$\mathbf{1}_{U}$ and its density matrix $\rho(U)$ is shown by that partition
and only that partition satisfying the:

\begin{center}
If $(  u,u^{\prime})  \in\operatorname*{indit}(
\mathbf{1}_{U})  $, then $u=u^{\prime}$

Partition logic Principle of Identity of Indistinguishables.
\end{center}

\noindent That is, if $u,u^{\prime}\in U$ are indistinguishable by the
discrete partition, i.e., $(  u,u^{\prime})  \in
\operatorname*{indit}(  \mathbf{1}_{U})  $, then they are
identical. This is trivial mathematically since $\operatorname*{indit}(
\mathbf{1}_{U})  =\Delta=\{  (  u,u)  :u\in U\}  $
and it comports with the fact that all \textit{other} partitions on $U$
contain at least one block with multiple elements, which may thus be interpreted as non-classical superposition events.

At this point, we may return to the question posed in the Introduction about
how does quantum notion of superposition differs from the classical
superposition of electromagnetic waves or even water waves. The ontic
difference is that in quantum superposition (as opposed to the classical
examples of wave superposition) is that the superposed definite- or
eigen-states are rendered indefinite on how they differ--which is variously
described in the literature as superpositions being blurry, unsharp, smudged,
blunt, coherent, fuzzy, blob-like, dispersed, smeared-out, indeterminate,
spread-out, or indefinite. In contrast, no such blurriness or indefiniteness
occurs in the classical superposition of, say, water or electromagnetic waves.
That is why the standard classroom ripple-tank model of the two-slit
experiment is seriously misleading since it represents superposition
classically as the addition of matter waves.

\begin{quotation}
\noindent Another surprising peculiarity of quantons is that they are blurry
or fuzzy rather than neat or sharp. Whereas in classical physics all
properties are sharp, in quantum physics only a few are: most are blunt or
smudged. ...

\noindent The reason for this fuzziness is that ordinarily an isolated quanton
is in a \textquotedblleft coherent\textquotedblright\ state, that is, the
combination or superposition (weighted sum) of two or more basic states (or
eigenfunctions). The superposition or \textquotedblleft
entanglement\textquotedblright\ of states is a hallmark of quantum mechanics.
\cite[pp. 49-50]{bunge:m-and-m}
\end{quotation}

This ontic difference comes out mathematically, not in the addition of the
$n\times1$ state vectors, but in the non-zero off-diagonal elements of the
$n\times n$ matrix treatment of density matrices (as well as in the prior
incidence matrix treatment of superposition events as opposed to classical
discrete events).

\section{Conclusion: The Born Rule}

The Born Rule does not occur in ordinary classical probability theory because
that theory does not include superposition events and the accompanying
amplitudes (that come from representing the density matrix of a superposition event as an outer product). When superposition events are introduced into the purely
mathematical theory, then the probability of outcomes can be computed as the
\textit{squares} of the coefficients in the normalized vector $\vert s \rangle$ associated with the superposition event $\Sigma S$. But
that means that the coefficients could be negative which is necessary if the
superposition events (as ``states'') are to
form a vector space. And the requirement that the state space be a vector
space is precisely what is required by the Superposition Principle as pointed
out by Dirac. 

\begin{quotation}
The superposition process is a kind of additive process and implies that
states can in some way be added to give new states. The states must therefore
be connected with mathematical quantities of a kind which can be added
together to give other quantities-of the same kind. The most obvious of such
quantities are vectors. ...

We now assume that each state of a dynamical system at a particular time
corresponds to a ket vector, the correspondence being such that if a state
results from the superposition of certain other states, its corresponding ket
vector is expressible linearly in terms of the corresponding ket vectors of
the other states, and conversely. \cite[pp. 15-16]{dirac:principles}
\end{quotation}

The pure density matrix $\rho(  \Sigma S)  $ can be constructed as
the outer product $\rho(  \Sigma S)  =\vert s\rangle
\langle s\vert $ where $\vert s\rangle $ is the $n$-ary
ket vector with the $i^{th}$ entry as the amplitude $\langle
u_{i}|s\rangle =\sqrt{\frac{p_{i}}{\Pr(  S)  }}\chi
_{S}(  u_{i})  =\sqrt{\frac{\Pr(  \{  u_{i}\}  \cap
S)  }{\Pr(  S)  }}$. Or starting with the \textit{pure}
density matrix $\rho(  \Sigma S)  $, then $\vert
s\rangle $ is obtained (up to sign) as the normalized eigenvector
associated with the eigenvalue of $1$ and $\rho(  \Sigma S)
=\vert s\rangle \langle s\vert $ is obtained as the
spectral decomposition of $\rho(  \Sigma S)  $ as a Hermitian matrix.

The probability of $u_{i}$ conditioned on the superposition event $\Sigma S$ is:

\begin{center}
$\Pr(  \{u_{i}\}|\Sigma S)  =\operatorname{tr}[  P_{\{
u_{i}\}  }\rho(  \Sigma S)  ]  =\frac{p_{i}}{\Pr(
S)  }\chi_{S}(  u_{i})  $.
\end{center}

\noindent The point is that this same probability conditioned by the
two-dimensional $n\times n$ density matrix $\rho(  \Sigma S)  $
could also be obtained from the ket vector $\vert s\rangle $ of
amplitudes as the square of the amplitudes:

\begin{center}
$\langle u_{i}|s\rangle ^{2}=\operatorname{tr}[  P_{\{
u_{i}\}  }\rho(  \Sigma S)  ]  =\frac{p_{i}}{\Pr(
S)  }\chi_{S}(  u_{i})  $.

The Born Rule (special case)
\end{center}

The Born Rule does not occur in classical finite probability theory since the
events $S$ are all discrete sets that can be represented by $n$-ary columns of
non-negative numbers. The associated $n\times n$ diagonal density matrix
$\rho(  S)  $ for the classical event $S$ is not the outer product
of a one-dimensional vector with itself (except when $S$ is a singleton, i.e.,
the null case of superposition). It has no non-zero off-diagonal elements
indicating the blurring or cohering together of the elements of $U$. Thus the
outcomes in a classical discrete event have probabilities, not amplitudes. To
accommodate the notion of a superposition event $\Sigma S$, it is necessary to
use two-dimensional $n\times n$ density matrices $\rho(  \Sigma S)
$ where the non-zero off-diagonal amplitudes indicate the blobbing or cohering
together in superposition of the elements associated with the corresponding
diagonal entries. And mathematically \textit{those} density matrices
$\rho(  \Sigma S)  $ (unlike $\rho(  S)  $) can be
constructed as the outer products $\vert s\rangle (
\vert s\rangle )  ^{t}=\vert s\rangle \langle
s\vert $ of ket vectors $\vert s\rangle $ of amplitudes. Then
the probability of the individual outcomes $\{  u_{i}\}  $
conditioned by the superposition event $\Sigma S$ is given as the
\textit{square} of amplitudes: $\langle u_{i}|s\rangle
^{2}=\operatorname{tr}[  P_{\{  u_{i}\}  }\rho(  \Sigma
S)  ]  =\frac{p_{i}}{\Pr(  S)  }\chi_{S}(
u_{i})  $. Of course, the probability of $u_{i}$ could also be obtained
as $\operatorname{tr}[  P_{\{  u_{i}\}  }\rho(  S)
]  $ but that method of encoding probabilities does not generalize to
vectors in a vector space, e.g., $\mathbb{R}^{n}$ or $\mathbb{C}^{n}$, with positive, negative, or complex components. Only the encoding of
probabilities as squared amplitudes has that feature which is the Born Rule.

Thus the Born Rule arises naturally out of the mathematics of probability
theory minimally enriched by superposition events and their associated
amplitudes. The Superposition Principle \cite[Chap. 1]{dirac:principles} in QM
requires that the states be represented as vectors within a full vector space
(not just non-negative vectors). The simple treatment of a superposition event
in the minimally enriched probability theory shows that the probabilities can
then be expressed as the squares of the normalized vector components (amplitudes). That
shows how probabilities can be generated with arbitrary normalized vectors,
e.g., with positive or negative components since their square is always positive.
  
In the Hilbert spaces over the complex numbers $\mathbb{C}$ of quantum
mechanics, the components in $\vert s\rangle $ may be complex (and the bra $\langle s \vert$ is the conjugate transpose of the corresponding ket) so
the square $\langle u_{i}|s\rangle ^{2}$ is then the
\textit{absolute} square $\vert \langle u_{i}|s\rangle
\vert ^{2}$. But that introduces nothing new in principle over what we
have shown here with real matrices arising from the extension of ordinary
probability theory with superposition events. 

Given the `mystery' that surrounds QM, it would perhaps be gratifying if the
Born Rule was some deep theorem (like the Spin-Statistics Theorem). But the
Born Rule does not need any more-exotic or physics-based explanation. Perhaps it is something of a `disappointment' that the Born
Rule is just a somewhat mundane feature of the mathematics of superposition. If an alleged ``derivation of the Born Rule'' does not start with probability theory, then the probabilities are likely ``put in by hand,'' i.e., postulated in some form. In contrast, we have shown how the Born Rule naturally arises by extending ordinary probability theory to superposition events in addition to the usual discrete events. Two-dimensional matrices are needed to differentiate superposition events from the ordinary discrete events, and those matrices can be obtained as the outer products of a vector of ``amplitudes'' and its transpose. In this approach, the extension of ordinary probability
theory to model superposition events shows how probabilities can be encoded using
(normalized) vectors of amplitudes as the (absolute) squares of those amplitudes. And superposition is, not coincidentally, the key \textit{non-classical} feature
of quantum mechanics as foreshadowed by the non-zero off-diagonal elements in
$\rho(  \Sigma S)  $ as opposed to $\rho(  S)  $. In short, the answer to the question: ``Where does the Born Rule come from?'' is superposition.

\section{Statements and Declarations}

The author did not receive support from any organization for the submitted
work. The author has no relevant financial or non-financial interests to
disclose. There are no conflicts of interest to disclose.

\end{document}